\documentclass[showpacs,showkeys,amssymb,amsmath,aps,twocolumn]{revtex4}
\usepackage{graphicx}
\usepackage{dcolumn}
\usepackage{bm}
\usepackage{epsfig}
\usepackage{color}
\begin{document}

\title{Modified strong-coupling treatment of a spin-1/2 Heisenberg trimerized chain developed from the exactly solved Ising-Heisenberg diamond chain}

\author{Taras Verkholyak}
\affiliation{Institute for Condensed Matter Physics, NASU, 1 Svientsitskii Street, L'viv-11, 79011, Ukraine}
\author{Jozef Stre\v{c}ka}
\affiliation{Department of Theoretical Physics and Astrophysics, Faculty of Science, P. J. \v{S}af\'{a}rik University, Park Angelinum 9, 040 01, Ko\v{s}ice, Slovakia}

\date{\today}

\begin{abstract}

Quantum spin-1/2 antiferromagnetic Heisenberg trimerized chain with strong intradimer and weak monomer-dimer coupling constants is studied using the novel many-body perturbation expansion, which is developed from the exactly solved spin-1/2 Ising-Heisenberg diamond chain preserving correlations between all interacting spins of the trimerized chain. It is evidenced that the novel perturbation approach is superior with respect to the standard perturbation scheme developed from a set of noninteracting spin monomers and dimers, whereas its accuracy even coincides up a moderate ratio of the coupling constants with the state-of-the-art numerical techniques. The Heisenberg trimerized chain shows the intermediate one-third plateau, which was also observed in the magnetization curve of the polymeric compound Cu$_3$(P$_2$O$_6$OH)$_2$ affording its experimental realization. Within the modified strong-coupling method we have obtained the effective Hamiltonians for the magnetic-field range from zero to one-third plateau, and from one-third plateau to the saturation magnetization. The unconventional second-order perturbation theory provides extremely accurate results for both critical fields of the intermediate one-third plateau up to the moderate ratio of the coupling constants as convincingly evidenced through a comparison with numerical density-matrix renormalization group data. It is shown that the derived effective Hamiltonian also provides at low enough temperatures sufficiently accurate results for magnetization curves and thermodynamic properties as corroborated through a comparison with quantum Monte Carlo simulations. Using the results for the effective Hamiltonian we additionally suggest straightforward procedure for finding the microscopic parameters of one-dimensional trimerized magnetic compounds with strong intradimer and weak monomer-dimer couplings. We found the refined values for the coupling constants of Cu$_3$(P$_2$O$_6$OH)$_2$ by matching the theoretical results with the available experimental data for the magnetization and magnetic susceptibility in a wide range of temperatures and magnetic fields.
\end{abstract}


\maketitle

\section{Introduction} 
\label{intro}

Low-dimensional quantum Heisenberg spin models show exceptionally diverse magnetic behaviour due to the competition between several factors: strong quantum fluctuations, breaking of full translational symmetry, frustrated interactions and external fields \cite{lnp645}. The most pronounced feature in such systems is the emergence of fractional plateaux in the magnetization curve \cite{b_takigawa11}. The possibility of observing the fractional magnetization plateaux in one-dimensional quantum spin models is governed by the Oshikawa-Yamanaka-Affleck rule \cite{oshikawa97}. In agreement with this rule, the one-third plateau was observed in the magnetization curve of the polymeric compound Cu$_3$(P$_2$O$_6$OH)$_2$, which represents an experimental realization of the spin-$\frac{1}{2}$ Heisenberg trimerized chain \cite{hase06,hase07,hase08,hase20,koo10,kong15}. To the best of our knowledge, the trimerized Heisenberg chain appeared as a first example for which the quantum magnetization plateau was theoretically predicted \cite{hida94}, whereas the difference between the quantum and classical fractional magnetization plateaux can be also analyzed on this paradigmatic example \cite{hida05}.

In view of its complexity, the quantum Heisenberg trimerized spin chain was studied mainly by numerical methods as for instance the exact diagonalization \cite{hida94}, the transfer-matrix renormalization group technique \cite{gu06} or the density-matrix renormalization group method \cite{gong08}. Exact solutions are available only for some limited version of the model as for instance $XY$ \cite{okamoto92,zvyagin91,ding10} and Ising-Heisenberg trimerized chains \cite{strecka03}. Among other analytical approaches one can mention series expansions for the plateau boundaries, which were found using the strong-coupling expansion from the limit of non-interacting dimers and trimers \cite{honecker99}. 

The asymmetric Heisenberg diamond chain represents a paradigmatic example of the trimerized spin chain with frustrating interactions, which is widely studied due to its relation to azurite (see e.g. Refs.~\cite{kikuchi05,jeschke11,honecker11}). The strong-coupling approach turned out to be a powerful tool for the study of low-temperature properties of the asymmetric Heisenberg diamond chain \cite{honecker01}. This approach takes into account the quantum corrections to the flat-band picture of magnon excitations inherent in the case of symmetric diamond chain and provides a relatively good description of the magnetization curve in the field range outside magnetization plateaux \cite{honecker01}. A further improvement of this approach was given in Ref. \cite{derzhko13}, where the localized-magnon approach was developed at high enough magnetic fields. 

Recently, the novel type of strong-coupling approach was developed from the exactly solved Ising-Heisenberg model on the symmetric diamond \cite{derzhko15} and orthogonal-dimer \cite{verkholyak16} chains. It has been verified that the modified calculation procedure shows a noticeable improvement with respect to the expansion around the limit of non-interacting dimers due to the fact that the unperturbed Hamiltonian already accounts for interdimer correlations between $z$-component of spins in addition to all intradimer correlations. Thus, the application of the strong-coupling expansion from the exact solution for the Ising-Heisenberg model seems to be quite a promising tool also for the trimerized antiferromagnetic Heisenberg chain,  which corresponds to the extremely asymmetric case of the diamond chain. The aim of this paper is twofold. First, we would like to use the quantum antiferromagnetic Heisenberg trimerized chain as a relatively simple example for the analysis of the novel perturbative expansion about the exact solution of the Ising-Heisenberg model. Second, we will show how the suggested approach can be implemented for the evaluation of the microscopic parameters of the polymeric magnetic compound Cu$_3$(P$_2$O$_6$OH)$_2$ \cite{hase06,hase07,hase08,hase20} serving as its experimental realization.

The plan of the paper is as follows. In Sec.~\ref{sec:eff_model}, we consider the perturbation theory about the Ising-Heisenberg limit and compare derived results with other numerical approaches. Here we also calculate thermodynamic quantities for the effective and original model in order to verify the validity of the established approach. All the numerical density-matrix renormalization group (DMRG) \cite{white92} and quantum Monte Carlo (QMC) \cite{sandvik99} simulations were performed by adapting routines from the ALPS project \cite{alps}.
In Sec.~\ref{sec:application} the results derived for the effective model are used to find the microscopic parameters of the polymeric compound Cu$_3$(P$_2$O$_6$OH)$_2$, which provide the appropriate description of its magnetic properties. The most important findings are summarized in Sec.~\ref{sec:concl}.

\section{Perturbation theory about the Ising-Heisenberg diamond chain}
\label{sec:eff_model}

\begin{figure}
\begin{center}
\includegraphics[width=0.5\textwidth]{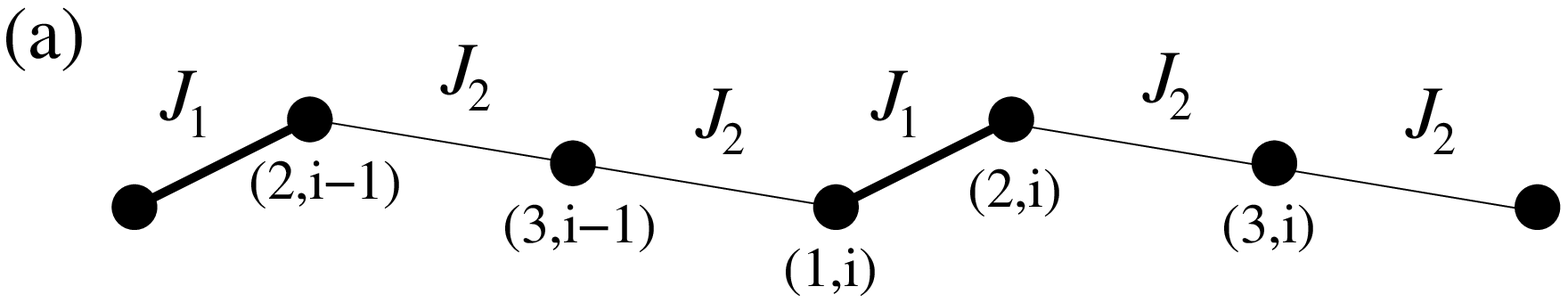}
\includegraphics[width=0.5\textwidth]{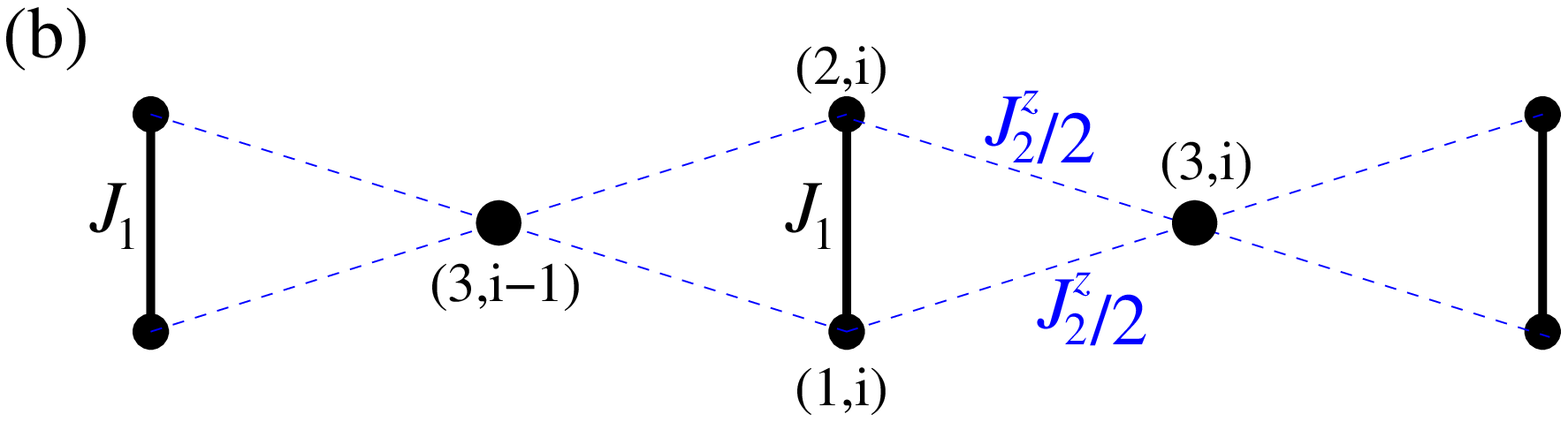}
\end{center}
\caption{(Color online) 
A schematic illustration of the spin-1/2 Heisenberg trimerized chain (1) (panel a) and the spin-1/2 Ising-Heisenberg diamond chain (2) (panel b) used for its effective description within the developed strong-coupling approach.}
\label{fig:model}
\end{figure}

Let us consider the quantum spin-$\frac{1}{2}$ Heisenberg trimerized chain (see Fig.~\ref{fig:model}(a)) described by the following Hamiltonian:
\begin{eqnarray}
H&=&\sum_{i=1}^{N}[J_1 {\mathbf s}_{1,i}\cdot{\mathbf s}_{2,i} 
+ J_2({\mathbf s}_{2,i}\cdot{\mathbf s}_{3,i} + {\mathbf s}_{3,i}\cdot{\mathbf s}_{1,i+1})
\nonumber\\
&&- h(s_{1,i}^z+s_{2,i}^z+s_{3,i}^z)],
\label{ham}
\end{eqnarray}
where ${\mathbf s}^{\alpha}_{m,i} = (s^{x}_{m,i}, s^{y}_{m,i}, s^{z}_{m,i})$ is the quantum spin-$\frac{1}{2}$ attached to the $m$-th site of the $i$-th unit cell, $J_1>0$ and $J_2>0$ are two different antiferromagnetic  interactions to be further referred to as the intradimer and monomer-dimer coupling constants, respectively, $h=g\mu_B B$ is the external magnetic field in energy units, $g$ is the Land\'e g-factor, $\mu_B$ is the Bohr magneton, 
$N$ is the number of unit cells, and the periodic boundary conditions are implied henceforth. 
The intradimer coupling constant $J_1$ is imposed to be substantially stronger than the monomer-dimer one $J_2$, i.e. $J_1 \gg J_2$, since we are interested in the regime of weakly coupled spin dimers and monomers. It is worthwhile to note that the opposite limit of weakly coupled spin trimers ($J_1 \ll J_2$) has been considered recently as well \cite{cheng21}.

The standard perturbation expansion would start from the limit of non-interacting dimers and monomers as for instance done in Ref. \cite{honecker01}. In the present work we would like to take into account at least some of the correlations between dimeric and monomeric spins quite rigorously. 
It is important to stress that the choice of the unperturbed Hamiltonian is not a trivial task. The initial guess, based on our preceding work on the orthogonal-dimer chain \cite{verkholyak16}, would be the spin-$\frac{1}{2}$ Ising-Heisenberg trimerized chain, where the weaker bonds $J_2$ are chosen to be of the Ising type \cite{strecka03}. However, this quantum spin-chain model exhibits a substantial zero plateau, which cannot be destroyed perturbatively by the $XY$-part of the monomer-dimer coupling constant $J_2$ and thus, it would show nonphysical behavior at sufficiently small magnetic fields. Hence, one has to search for such a model among the class of exactly solvable spin chains that is reminiscent of basic properties of the Heisenberg trimerized chain (\ref{ham}). In this regard, the spin-$\frac{1}{2}$ Ising-Heisenberg diamond chain \cite{canova06} with the macroscopically degenerate ground state in zero field seems to be more appropriate for a development of the many-body perturbation theory, which takes into account spin correlations between $z$-components quite rigorously. In what follows we will therefore divide the initial Hamiltonian (\ref{ham}) into two parts $H=H_{IH} + V$, whereas the unperturbed part $H_{IH}$ corresponds to the exactly solved spin-$\frac{1}{2}$ Ising-Heisenberg diamond chain (see Fig.~\ref{fig:model}(b)) \cite{canova06}:
\begin{eqnarray}
\label{ham_IH}
H_{IH}&=&\sum_{i=1}^N \left[ J_1{\mathbf s}_{1,i}\cdot{\mathbf s}_{2,i} + \frac{J^z_2}{2}(s^z_{1,i} + s^z_{2,i})(s^z_{3,i-1}+s^z_{3,i})
\right.\nonumber\\ 
&& \left.
-h'\sum_{m=1}^3 s^z_{m,i}\right],
\end{eqnarray}
and the perturbed part $V$ contains the rest of the total Hamiltonian (\ref{ham}):
\begin{eqnarray}
\label{ham_pert}
V&=&\sum_{i=1}^N \left[  
J^{xy}_2\sum_{\alpha=x,y}(s^\alpha_{2,i}s^\alpha_{3,i} + s^\alpha_{3,i}s^\alpha_{1,i+1})
\right.
\nonumber\\
&{+}&\left.
\frac{J^z_2}{2}(s^z_{1,i} {-} s^z_{2,i})(s^z_{3,i-1} {-} s^z_{3,i})
{-}(h{-}h')\sum_{m=1}^3 s^z_{m,i} 
\right].
\end{eqnarray}
In above, the parameter $h'$ denotes the critical field of the Ising-Heisenberg diamond chain at which the ground state changes. It will be defined separately for two particular cases with the ground-state macroscopic degeneracy emergent either at zero or saturation field. Note furthermore that we  have formally distinguished the $xy$- and $z$-components of the monomer-dimer coupling constant, which are ultimately set equal to each other $J_2^{xy} = J_2^z = J_2$ in order to recover the isotropic nature of this exchange interaction. The ground state of the spin-$\frac{1}{2}$ Ising-Heisenberg diamond chain (\ref{ham_IH}) is rather simple for $J_2<J_1$ (see Ref. \cite{canova06} and the derivation in Appendix \ref{app:IH-dc}), because it only contains two ground states referred to as the monomer-dimer (MD) phase (\ref{MD}) and the saturated paramagnetic (SAT) phase (\ref{SAT}) coexisting together at the saturation field $h_c = J_1+J^z_2/2$.

In the following we will adapt the standard many-body perturbation scheme for the Hamiltonian $H=H_0 + V$, which can be recast into the perturbed part $V$ and the unperturbed part $H_0$ with rigorously known eigenenergies $E_i^{(0)}$ and eigenvectors $|\Phi^{(0)}_i\rangle$ derived from the eigenvalue equation $H_0|\Phi^{(0)}_i\rangle=E_i^{(0)}|\Phi^{(0)}_i\rangle$ \cite{b_fulde91}. 
If $P$ denotes the projection operator on the subspace of states corresponding to the lowest eigenstate $E_i^{(0)}$ of the unperturbed model given by the Hamiltonian $H_0$ and $Q=1-P$, then, the perturbative expansion can be formally written for the effective Hamiltonian acting in this subspace:
\begin{eqnarray}
&&H_{eff}=PHP + PV R_s\sum_{n=0}^{\infty}
\left[ (QV - \delta E_0) R_s\right]^n QVP, 
\nonumber\\
&& R_s=Q\frac{1}{E_0^{(0)}-H_0}=\sum_{m\neq 0}\frac{|\Phi^{(0)}_m\rangle\langle\Phi^{(0)}_m|}{E_0^{(0)}-E_m^{(0)}},
\label{eq_pert}
\end{eqnarray}
where $\delta E_0=E_0-E_0^{(0)}$. 

In the present work we will restrict ourselves to the second-order perturbative expansion, which will take into account the zeroth-, first- and the second-order contributions to the effective Hamiltonian: $H_{eff}^{(0)}=PH_0P$, $H_{eff}^{(1)}=PVP$ and $H_{eff}^{(2)}=PV R_s VP$, respectively. The main originality and novelty of the developed perturbation method lies in the choice of the unperturbed part of the Hamiltonian $H_0$, which is identified with the Hamiltonian $H_{IH}$ of the spin-$\frac{1}{2}$ Ising-Heisenberg diamond chain defined by Eq. (\ref{ham_IH}). It is evident that the unperturbed Hamiltonian $H_{IH}$ pertinent to this exactly solved model preserves correlations between all interacting spins of the quantum spin chain in opposite to the unperturbed Hamiltonian corresponding to a set of noninteracting monomers and dimers being the starting ground for the standard perturbation theory.

At sufficiently low magnetic fields the ground state of the spin-$\frac{1}{2}$ Ising-Heisenberg diamond chain is the MD phase, which becomes macroscopically degenerate in zero field due to a paramagnetic (free) character of the monomeric spins ${\mathbf s}_{3,i}$ effectively decoupled by singlet states of the dimeric spins ${\mathbf s}_{1,i} \mbox{---} {\mathbf s}_{2,i}$ (see Appendix~\ref{app:IH-dc}). Of course, any nonzero magnetic field lifts the macroscopic degeneracy of the MD phase, which will be restored just at the saturation field $h_c$ where the dimeric spins ${\mathbf s}_{1,i} \mbox{---} {\mathbf s}_{2,i}$ undergo a change from the singlet-dimer to the polarized triplet state. Owing to this fact, we have to distinguish two particular cases requiring separate considerations: a) small enough magnetic fields $h \ll h_c$; and b) higher magnetic fields sufficiently close to the saturation field $h_c$.

\subsection{Low-field region, $0\leq m\leq 1/3$}
\label{subs:small_fields}

The perturbative treatment of $V$ up to the second order results in the following effective Hamiltonian (see Appendix~\ref{app:small_fields} for more details): 
\begin{eqnarray}
\label{hef_a}
H_{eff}&=&\sum_{i=1}^{N}\left(
J_{eff}{\mathbf s}_{3,i}\cdot{\mathbf s}_{3,i+1} -h {s}_{3,i}^z
\right), 
\nonumber\\
J_{eff}&=&\frac{J_2^2}{2J_1-J_2}>0.
\end{eqnarray}
At sufficiently low fields the spin-$\frac{1}{2}$ Heisenberg trimerized chain can be thus effectively described by the antiferromagnetic spin-$\frac{1}{2}$ Heisenberg chain with the effective coupling constant $J_{eff}$ given by Eq. \eqref{hef_a}. The saturation field of the effective spin-$\frac{1}{2}$ Heisenberg chain given by the Hamiltonian \eqref{hef_a} is known exactly and it corresponds to the lower critical field of the 1/3-plateau phase 
\begin{eqnarray}
\label{hef_al}
h_{1/3}^{l}=2 J_{eff}=\frac{2J_2^2}{2J_1-J_2}, 
\end{eqnarray}
at which all monomeric spins ${\mathbf s}_{3,i}$ become fully polarized in the magnetic field.

\subsection{High-field region, $1/3\leq m \leq m_{sat}$}
\label{subs:strong_fields}
Next, we will consider the high-field region where the magnetization changes from 1/3-plateau to the saturation value. At the saturation field $h_c$ the spin-$\frac{1}{2}$ Heisenberg trimerized chain is macroscopically degenerate because the dimeric spins ${\mathbf s}_{1,i} \mbox{---} {\mathbf s}_{2,i}$ may be either in singlet or polarized triplet state. In this particular case, the application of the many-body perturbation theory leads to the effective spin-$\frac{1}{2}$ $XXZ$ Heisenberg chain (see Appendix~\ref{app:high_fields} for more details):
\begin{eqnarray}
\label{hef_b}
H_{eff}^{xxz}&=&\sum_{i=1}^{N/2}\left[
J^z_{eff}\tilde{s}_i^z \tilde{s}_{i+1}^z
+J^{xy}_{eff} (\tilde{s}_i^x \tilde{s}_{i+1}^x + \tilde{s}_i^y \tilde{s}_{i+1}^y) 
-\tilde{h} \tilde{s}_i^z
\right], 
\nonumber\\
J^{xy}_{eff}&=&\frac{J_2^2}{2 (2J_1 - J_2)}>0,
\nonumber\\
J^z_{eff}&=&\frac{J_2}{2J_1}J^{xy}_{eff},
\nonumber\\
\tilde{h}&=&h-J_1-\frac{J_2}{2}-\frac{J_2^2}{4J_1}.
\end{eqnarray}
Note that the quasi-spin operators $\tilde{s}^\alpha_i$ act in the space, where $|\tilde\downarrow\rangle_i$ ($|\tilde\uparrow\rangle_i$) corresponds to the singlet state $|0\rangle_i$ (the polarized triplet state $|1\rangle_i$) of the $i$-th dimeric unit ${\mathbf s}_{1,i} - {\mathbf s}_{2,i}$. The ground-state magnetization of the spin-$\frac{1}{2}$ Heisenberg trimerized chain can be expressed in terms of new operators according to the following formula:
\begin{eqnarray}
m=\frac{1}{3N}\sum_{i=1}^N\sum_{m=1}^3\langle s^z_{m,i} \rangle
=\frac{1}{3N}\sum_{i=1}^N(1+\langle \tilde{s}^z_{i} \rangle).
\end{eqnarray}
The critical fields of the effective spin-$\frac{1}{2}$ $XXZ$ Heisenberg chain are known exactly to be 
$\tilde{h}_{\pm}=\pm(J^z_{eff}+J^{xx}_{eff})$. Thus, one gets the following results for the saturation field and the upper critical field of the 1/3-plateau phase: 
\begin{eqnarray}
\label{cf}
&&h_{SAT}=J_1+\frac{J_2}{2}+\frac{J_2^2}{2J_1-J_2},
\nonumber\\
&&h_{1/3}^u
=J_1 + \frac{J_2}{2} - \frac{J_2^3}{2J_1(2J_1-J_2)}.
\end{eqnarray}

\subsection{Comparison with other theories}
\label{subs:comparison}

\begin{figure}
\begin{center}
\includegraphics[width=0.5\textwidth]{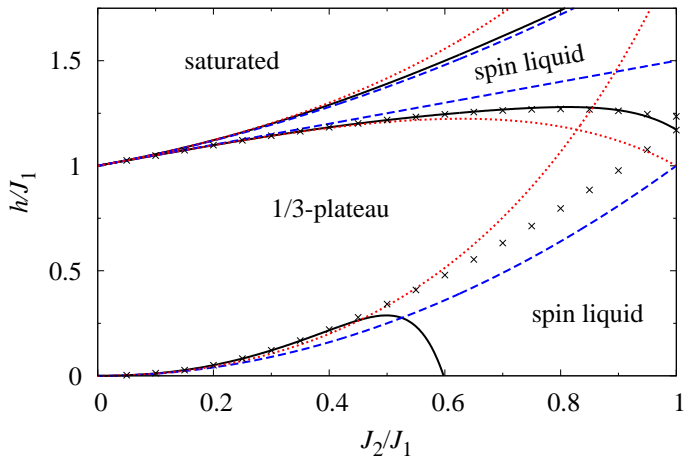}
\end{center}
\caption{(Color online) The ground-state phase diagram of the spin-$\frac{1}{2}$ Heisenberg trimerized chain. Solid lines correspond to the series expansion adapted according to Ref. \cite{honecker99}, dashed lines show the second-order strong-coupling approach developed from the monomer-dimer limit \cite{honecker01,honecker11}, dotted lines correspond to the second-order perturbation theory developed from the spin-$\frac{1}{2}$ Ising-Heisenberg diamond chain (present work), whereas the symbols represent the DMRG simulations of the original Heisenberg trimerized spin chain (\ref{ham}) with the number of cells $L=28$ (the number of kept states was 800). 
The tiny 1/3 plateau for the DMRG simulations at $J_2/J_1=1$ is the result of the finite size effect, and it obviously disappears in the thermodynamic limit.}
\label{fig:phase_diag}
\end{figure}

The spin-$\frac{1}{2}$ Heisenberg distorted diamond chain and the spin-$\frac{1}{2}$ Heisenberg trimerized chain as its special limiting case were previously studied in Refs. \cite{honecker99,honecker01,derzhko13}. 
In case $J_2<J_1$ it exhibits two gapped phases corresponding to the 1/3 plateau and the saturation, and the gapless quantum spin liquid phases where the magnetization changes continuously in a field (see Fig.~\ref{fig:phase_diag}).
The ground-state phase diagram shown in Fig.~\ref{fig:phase_diag} compares the critical fields obtained within four different calculation schemes: the series expansion (solid lines) \cite{honecker99}, the second-order strong-coupling approach developed from the monomer-dimer limit (dashed lines) \cite{honecker01,honecker11}, the second-order perturbation scheme developed in the present work from the spin-$\frac{1}{2}$ Ising-Heisenberg diamond chain (dotted lines) and the numerical DMRG simulations (symbols). The series expansion for the critical fields \cite{honecker99} is very accurate except the lower critical field $h_{1/3}^l$ of the 1/3-plateau phase, which starts to considerably decay from the most accurate DMRG data above $J_2/J_1\gtrsim 0.5$. Surprisingly, the second-order perturbation theory developed from the exactly solved spin-$\frac{1}{2}$ Ising-Heisenberg diamond chain also produces quite precise results for the critical fields in a wide range of the parameter space $J_2/J_1 \lesssim 0.6$. With exception of the saturation field, the present perturbation theory provides significantly better results for the critical fields in comparison with the strong-coupling approach developed up to the second order from the limit of non-interacting monomers and dimers \cite{honecker01,honecker11}. This simpler strong-coupling approach provides for sufficiently small fields the effective spin-$\frac{1}{2}$ Heisenberg chain with the isotropic coupling constant $J_{eff}=\frac{J_2^2}{2 J_1}$ (see Eq.~(5.6) in Ref.~\cite{honecker01}), while for high enough fields it provides the effective spin-$\frac{1}{2}$ XX chain with the coupling constant $J_{xy}=\frac{J_2^2}{4J_1}$ (see Eqs.~(7)-(10) in Ref.~\cite{honecker11}). 
It is therefore of particular interest to compare the precision of both perturbative methods: the aforementioned results coincide with our results in the second order with respect to the coupling ratio $J_2/J_1$. As far as higher values of the interaction ratio $J_2/J_1$ are concerned, the 1/3 plateau disappears within the presented unconventional perturbative scheme at $J_2/J_1\approx 0.8$, while the simpler strong-coupling method preserves this plateau until $J_2/J_1=1$.
It is also worthwhile to remark that the effective model \eqref{hef_b} and the respective critical fields \eqref{cf} derived in the high-field regime coincide with the results of the localized-magnon approach when considering the limit of trimerized chain \cite{derzhko13}. 

\begin{figure}
\begin{center}
\includegraphics[width=0.5\textwidth]{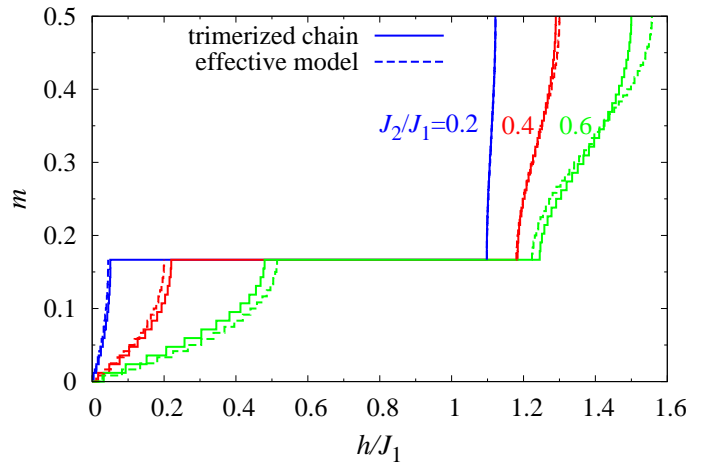}
\end{center}
\caption{(Color online) The zero-temperature magnetization curves of the spin-$\frac{1}{2}$ Heisenberg trimerized chain for three different values of the interaction ratio $J_2/J_1$ = 0.2, 0.4 and 0.6. 
Solid lines correspond to DMRG simulations of the original Heisenberg trimerized spin chain (\ref{ham}), while dashed lines were calculated with the help of the effective models (\ref{hef_a}) and (\ref{hef_b}).
}
\label{fig:magnet_T0}
\end{figure}

Zero-temperature magnetization curves obtained from the DMRG simulations of the original spin-$\frac{1}{2}$ Heisenberg trimerized chain (\ref{ham}) are compared in Fig.~\ref{fig:magnet_T0} with analogous magnetization data obtained with the help of the effective models (\ref{hef_a}) and (\ref{hef_b}) being valid in the low- and high-field range, respectively. It can be seen from Fig.~\ref{fig:magnet_T0} that the magnetization data obtained from the effective models show a reliable agreement with the magnetization curves of the spin-$\frac{1}{2}$ Heisenberg trimerized chain even up to relatively large values of the interaction ratio $J_2/J_1 \approx 0.6$, whereas the largest discrepancy can be detected in a vicinity of the saturation field.

\subsection{Thermodynamic properties}
\label{sec:thermodynamics}

\begin{figure}
\includegraphics[width=0.48\textwidth]{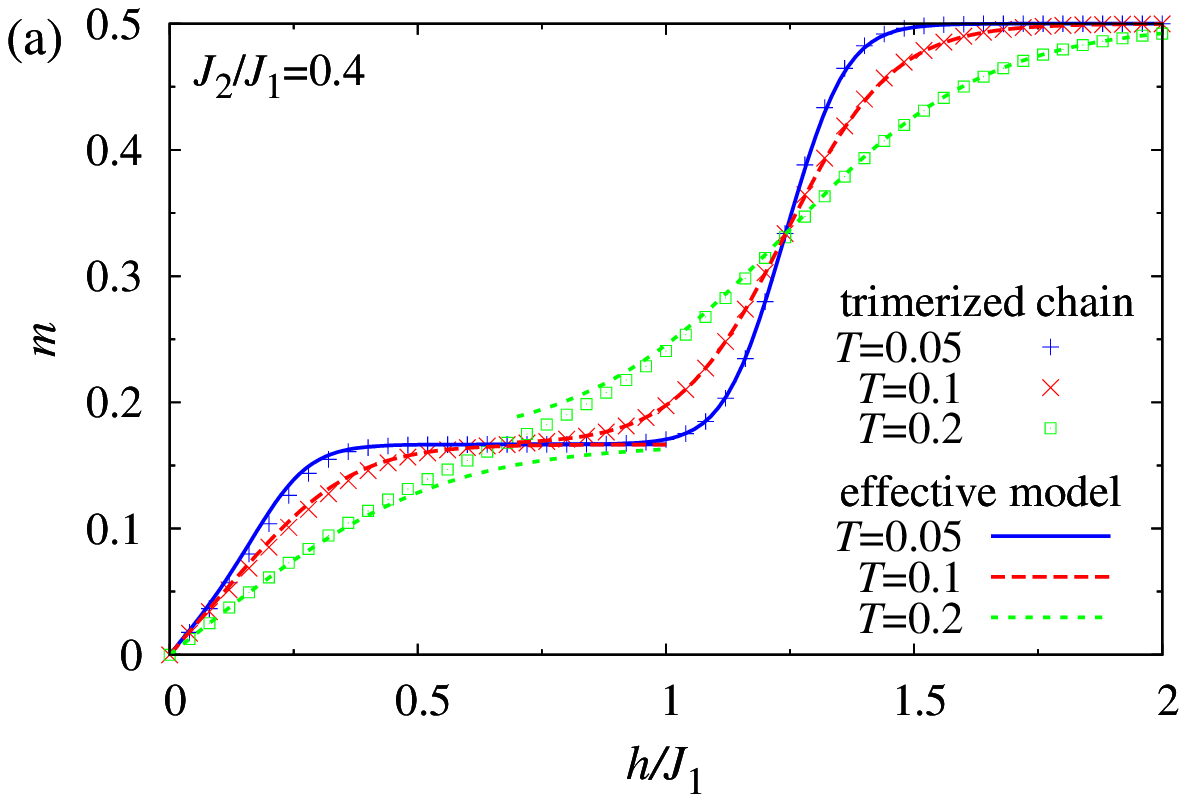}
\includegraphics[width=0.48\textwidth]{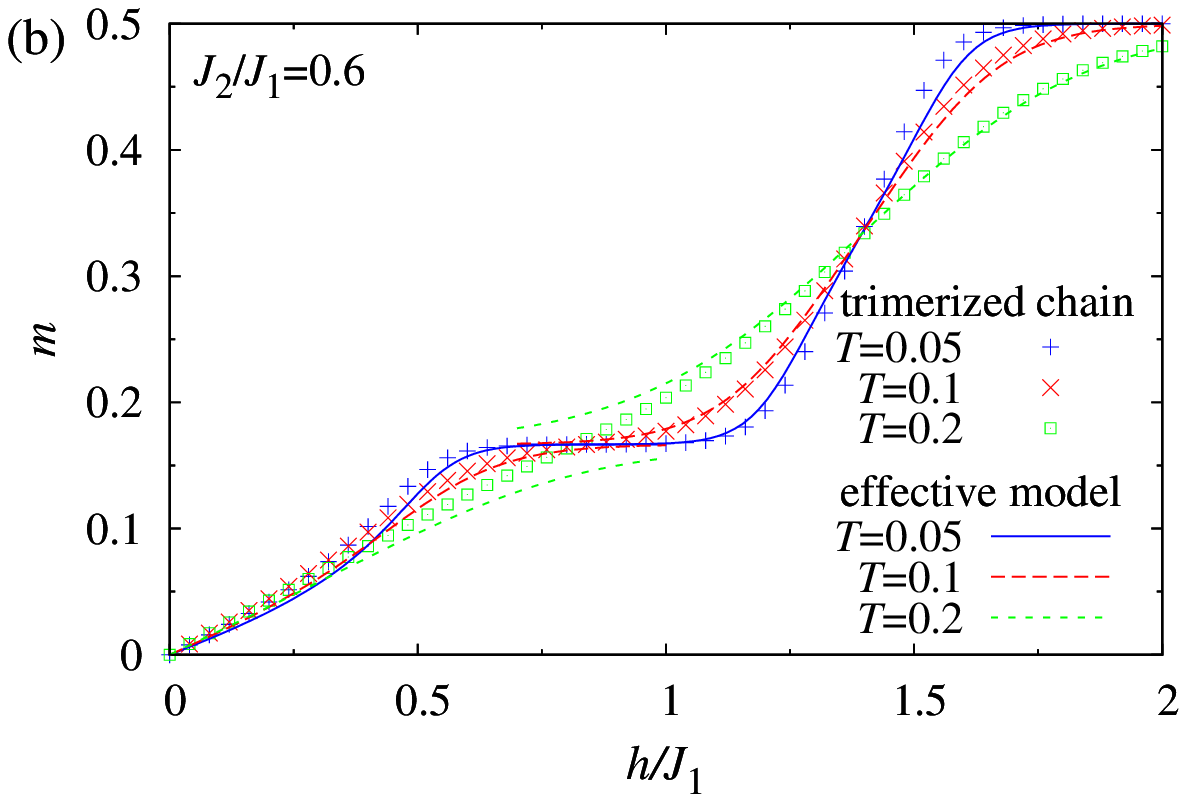}
\caption{(Color online) A comparison between the magnetization curves of the spin-$\frac{1}{2}$ Heisenberg trimerized chain obtained by using QMC simulations of the original model (\ref{ham}) and the effective models (\ref{hef_a}) and (\ref{hef_b}) at three different temperatures $k_B T/J_1$ = 0.05, 0.1 and 0.2 for $J_2/J_1=0.4$ (a), $J_2/J_1=0.6$ (b). QMC simulations were performed for the number of unit cells $L=120$, whereas $5 \times 10^5$ QMC steps were used for thermalization and additional $5 \times 10^6$ QMC steps for statistical averaging.
}
\label{fig:magnet}
\end{figure}

Although the main accent of a description based on the effective Hamiltonians (\ref{hef_a}) and (\ref{hef_b}) lies on a simpler ground-state analysis, the effective models are capable of describing appropriately the low-temperature thermodynamics as well. The validity of this perturbative approach depends on the energy gap between the ground and excited states of the unperturbed model so that the temperature range for its applicability is restricted to $T\ll\Delta E=J_1 \pm (\frac{J_2}{2}-h)$. In the following, we will examine in detail to what extent the effective models given by the Hamiltonians (\ref{hef_a}) and (\ref{hef_b}) are capable of describing thermodynamic properties of the spin-$\frac{1}{2}$ Heisenberg trimerized chain at nonzero temperatures with reasonable accuracy. For this purpose, we have calculated a few isothermal magnetization curves of the spin-$\frac{1}{2}$ Heisenberg trimerized chain using the QMC simulations of the original model (\ref{ham}) and the effective models (\ref{hef_a}) and (\ref{hef_b}), respectively. The results of QMC simulations presented in Fig.~\ref{fig:magnet} show a good agreement between the magnetization data of the original and effective models up to temperature $T/J_1 \lesssim 0.2$ below which excited states with higher energies ignored in the effective Hamiltonians (\ref{hef_a}) and (\ref{hef_b})  are not as much important. 

\begin{figure}
\includegraphics[width=0.48\textwidth]{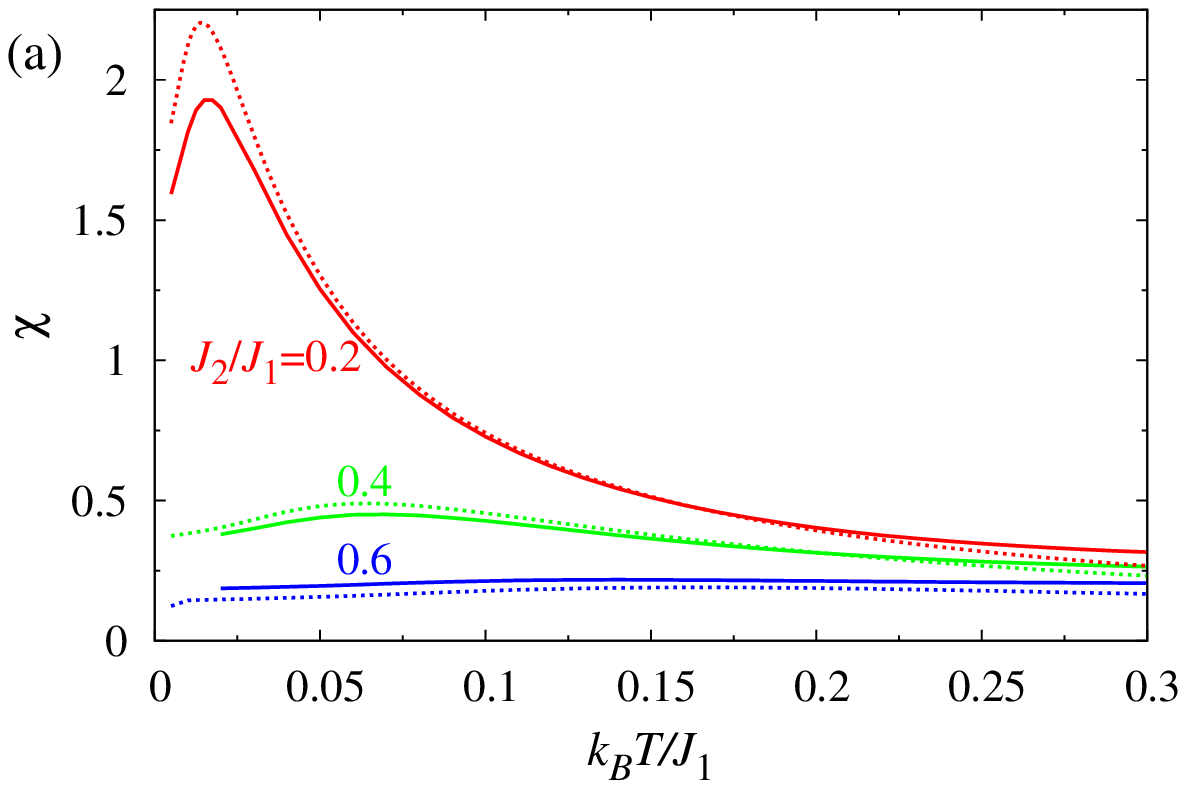}
\includegraphics[width=0.48\textwidth]{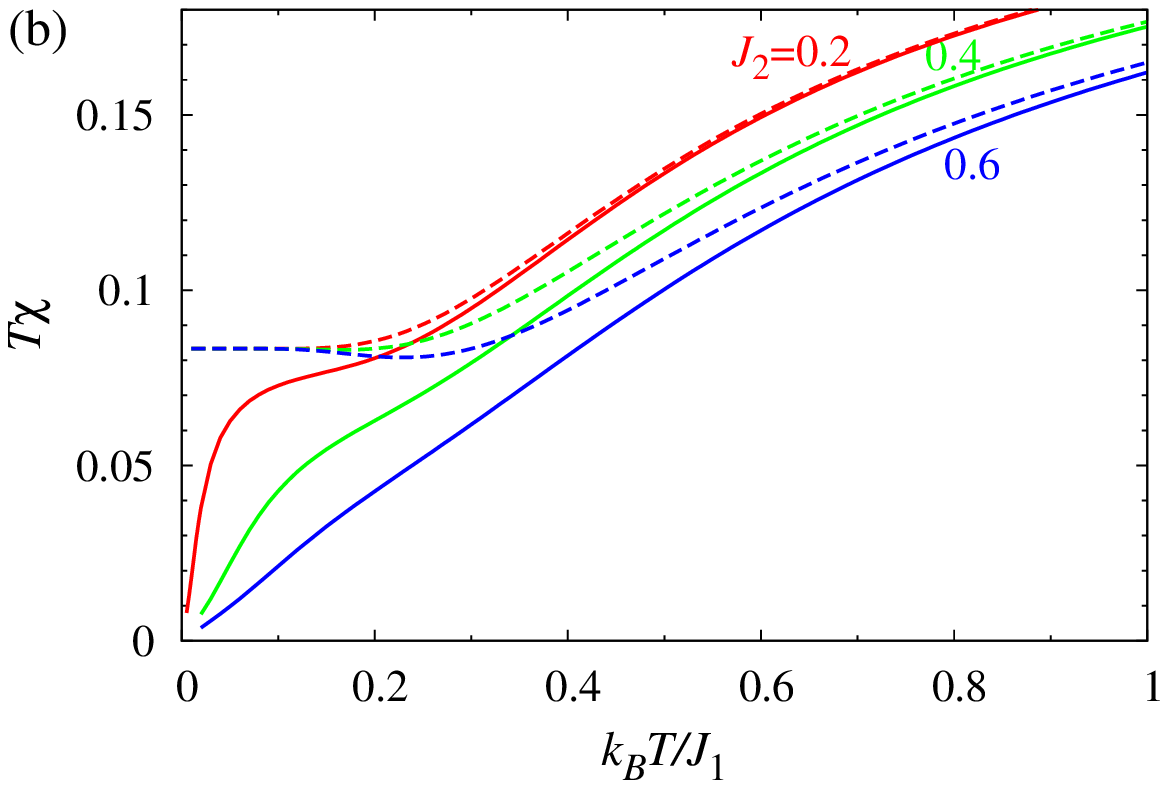}
\caption{(Color online) Temperature variations of the zero-field susceptibility $\chi$ (a) and the zero-field susceptibility times temperature product $\chi T$ (b) of the spin-$\frac{1}{2}$ Heisenberg trimerized chain for three different values of the interaction ratio $J_2/J_1 = 0.2$, 0.4 and 0.6. QMC simulations were performed for the number of unit cells $L=120$, whereas $10^7$ QMC steps were used for thermalization and additional $2 \times 10^7$ QMC steps for statistical averaging. Solid lines correspond to the original model (\ref{ham}), dotted lines correspond to the effective model (\ref{hef_a}) and dashed lines come from first-order linked cluster expansion given by Eq. \eqref{chi_int-mon-dim}.}
\label{fig:suscept-t}
\end{figure}
\begin{figure}
\includegraphics[width=0.48\textwidth]{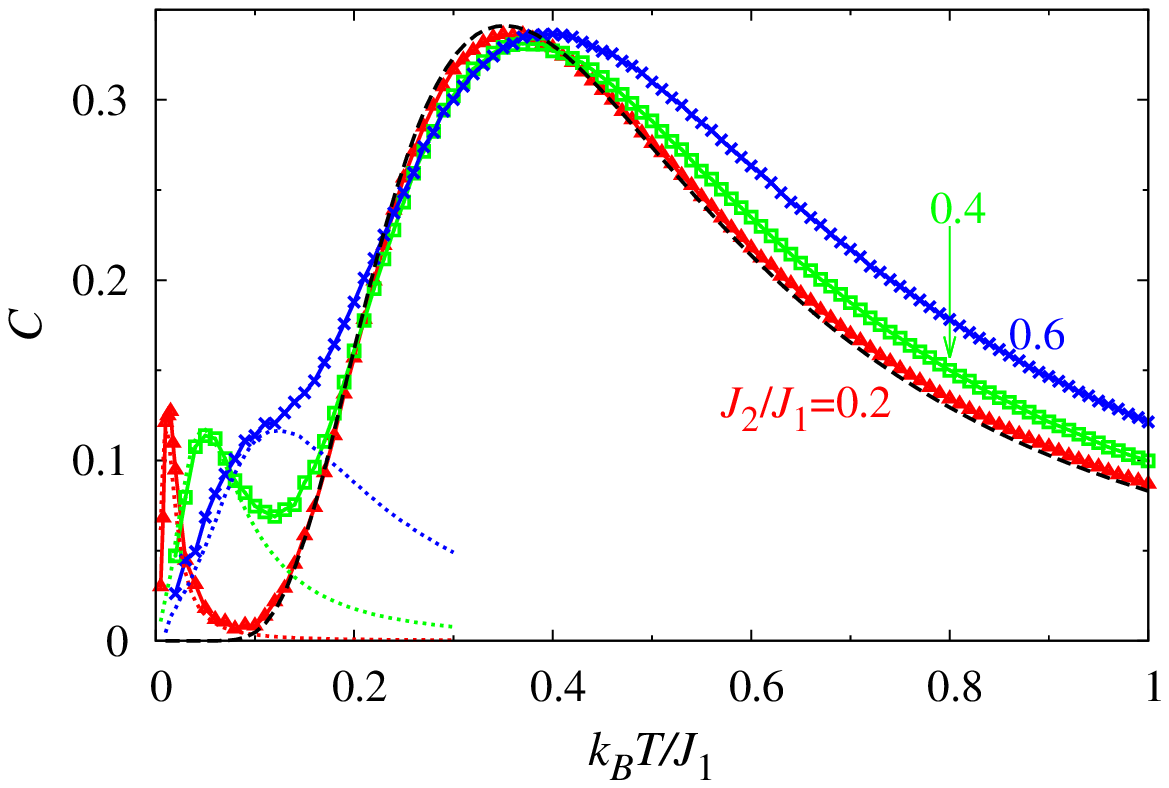}
\caption{(Color online) Temperature dependence of the zero-field specific heat of the spin-$\frac{1}{2}$ Heisenberg trimerized chain for three different values of the interaction ratio $J_2/J_1 = 0.2$, 0.4 and 0.6. QMC simulations were performed for the number of unit cells $L=120$, whereas $10^7$ QMC steps were used for thermalization and additional $2 \times 10^7$ QMC steps for statistical averaging. Solid lines correspond to the original model (\ref{ham}), dotted lines correspond to the effective model (\ref{hef_a}) and a dashed line indicates the contribution of isolated spin dimers \cite{haraldsen05} as given by Eq. \eqref{heat}.}
\label{fig:heat-t}
\end{figure}
Fig.~\ref{fig:suscept-t}(a) compares the zero-field susceptibility of the original (\ref{ham}) and effective (\ref{hef_a}) models at sufficiently low temperatures. It is evident that the zero-field susceptibility of the effective model (\ref{hef_a}) exhibits a low-temperature maximum at the same temperature as the initial model (\ref{ham}), whereas its height is just slightly overestimated. Similarly, the low-temperature peak of the specific heat can be also reasonably approximated by the effective model (\ref{hef_a}), see Fig.~\ref{fig:heat-t}. The low-temperature peak of the specific heat is indeed satisfactorily reproduced not only for sufficiently weak coupling constant between the monomeric and dimeric spins $J_2/J_1=0.2$, but also for its moderate value $J_2/J_1=0.4$. The effective model (\ref{hef_a}) provides a reasonable estimate of the low-temperature specific heat even for relatively strong value of the interaction ratio $J_2/J_1=0.6$ when the low-temperature maximum is superimposed in the form of shoulder on a round maximum originating from the intradimer excitations. 

In addition, we have also performed thermodynamic perturbation theory starting from the limit of noninteracting dimers and monomers to find the asymptotic behavior of thermodynamic quantities at sufficiently high temperatures. To this end, we  have divided the total Hamiltonian into two parts, the part involving the noninteracting dimers and monomers in a magnetic field:
\begin{eqnarray}
H_{md}^{(0)}=\sum_{i=1}^{N}[J_1 {\mathbf s}_{1,i}\cdot{\mathbf s}_{2,i} 
- h(s_{1,i}^z+s_{2,i}^z+s_{3,i}^z)],
\label{ham_MD0}
\end{eqnarray}
and the part incorporating the coupling $J_2$ between the monomeric and dimeric spins:
\begin{eqnarray}
H_{md}^{(1)}=\sum_{i=1}^{N}[
 J_2({\mathbf s}_{2,i}\cdot{\mathbf s}_{3,i} + {\mathbf s}_{3,i}\cdot{\mathbf s}_{1,i+1})].
\label{ham_MD00}
\end{eqnarray}
Then, the free energy can be found using the linked cluster expansion (see e.g. Ref.~\cite{mahan}):
\begin{eqnarray}
F&=&F_0-\frac{1}{\beta}\ln\langle\sigma(\beta)\rangle_0,
\label{free_en}
\\
\sigma(\beta)&=&T_\tau\exp\left(\int_0^{\beta}d\tau H_{md}^{(1)}(\tau)\right),
\nonumber
\end{eqnarray}
where $\beta = 1/(k_B T)$, $k_B$ is Boltzmann's constant, $T$ is absolute temperature, $\langle\cdots\rangle_0$ denotes the thermodynamic averaging over the decoupled monomer-dimer model \eqref{ham_MD0} and $H_{md}^{(1)}(\tau)=\exp(\tau H_{md}^{(0)}) H_{md}^{(1)} \exp(-\tau H_{md}^{(0)})$. Within the first-order expansion, we have obtained the following expression for the zero-field susceptibility:
\begin{eqnarray}
\label{chi_int-mon-dim}
\chi = \frac{g^2\mu_B^2 \beta}{3}
\left[\frac{1}{4} {+} \frac{2}{3+\exp(\beta J_1)} {-} \frac{\beta J_2}{3{+}\exp(\beta J_1)} \right],
\end{eqnarray}
where $g$ is gyromagnetic factor and $\mu_B$ is Bohr magneton. Analogously, the specific heat in the first-order linked cluster expansion coincides with the specific heat of the decoupled model \cite{haraldsen05}:
\begin{eqnarray}
\label{heat}
C=\frac{3\beta^2 J_1^2 \exp(-\beta J_1)}{[1+3\exp(-\beta J_1)]^2}.
\end{eqnarray}

The results  \eqref{chi_int-mon-dim} and \eqref{heat} of the linked cluster expansion for the susceptibility and specific heat are shown in Figs.~\ref{fig:suscept-t}(b) and \ref{fig:heat-t} together with the QMC data. The susceptibility (\ref{chi_int-mon-dim}) closely follows the QMC results for the trimerized chain down to  moderate temperatures $k_B T / J_1 \approx 0.5$. On the other hand, the round high-temperature maximum of the specific heat perfectly coincides with the dimer contribution (\ref{heat}) for small enough values of the interaction ratio $J_2/J_1$.

\section{Theory versus experiment}
\label{sec:application}

The magnetic compound Cu$_3$(P$_2$O$_6$OH)$_2$, which represents an experimental realization of the spin-$\frac{1}{2}$ Heisenberg trimerized chain \cite{hase06,hase07,hase08}, affords a suitable testbed for verification of the efficiency of the developed perturbation theory. Before doing so, let us summarize here the main  outcomes reported in the previous experimental studies \cite{hase06,hase07,hase08}. The high-field magnetization measurements on the magnetic compound Cu$_3$(P$_2$O$_6$OH)$_2$ revealed presence of the intermediate 1/3-plateau, which is detected in a low-temperature magnetization curve above 12~T. A temperature dependence of the magnetic susceptibility displays a round maximum with the peak height $\chi_{max}=0.0154$ emergent at $T_{max}\approx 3.25$~K \cite{hase06}, while the gyromagnetic $g$-factor was determined by ESR measurements as $g = 2.12$ \cite{hase06}. The distinct inelastic peak at $\omega=9.8$~meV$=113.7$~K, which is only weakly dependent on the wave vector, was detected by inelastic neutron scattering \cite{hase07,hase08}.

First, we will exemplify how the susceptibility data  measured in a wide range of temperatures may provide sufficient information for the evaluation of the microscopic parameters of the spin-$\frac{1}{2}$ Heisenberg trimerized chain. It is worthwhile to recall that the susceptibility of the spin-$\frac{1}{2}$ Heisenberg chain with the coupling constant $J_{eff}$ has the peak $\chi_{max}^{H}\approx 0.147/J_{eff}$ at $k_B T_{max}^{H}/J_{eff}\approx 0.640824$ \cite{eggert94}. The similar finding was also reported for the maximum of the specific heat: $C_{max}^{H}\approx 0.35/J_{eff}$ at  $k_B T_{max}^{c}/J_{eff}\approx 0.481$ \cite{bonner64}. Since the spin-$\frac{1}{2}$ Heisenberg trimerized chain can be faithfully represented at low enough temperatures by the effective spin-$\frac{1}{2}$ Heisenberg chain (\ref{hef_a}) one may immediately find the effective coupling $J_{eff}$ when comparing the results of Ref. \cite{eggert94} with the position of susceptibility maximum observed in experiment:
\begin{eqnarray}
J_{eff}/k_B\approx T_{max}/0.640824=5.071595321~{\rm K}.
\label{j_eff}
\end{eqnarray}
According to Eq.~(\ref{hef_a}), the effective coupling $J_{eff}/k_B \approx 5.071$ provides a useful relation connecting the intradimer $J_1$ and dimer-monomer $J_2$ coupling constants. Another independent relation connecting both coupling constants can be obtained from the high-temperature behavior of the susceptibility times temperature product $\chi T$. At very high temperatures the paramagnetic behavior prevails, and, thus, the product $\chi T$ tends to the Curie constant $C=\frac{N g^2\mu_B^2 S(S+1)}{3k_B}$, where $S$ is the spin of the magnetic unit. However, the sizable exchange coupling may have a significant effect even at relatively high temperatures $T\sim 200-300$~K, where the product $\chi T$ seems to be still far from being saturated. To get the appropriate description of the high-temperature susceptibility one may take advantage of the first-order linked cluster expansion within the thermodynamic perturbation theory as given by Eq.~(\ref{chi_int-mon-dim}). Implementing the nonlinear least-squares (NLLS) Marquardt-Levenberg algorithm implemented in gnuplot \cite{gnuplot} for the fitting of the susceptibility data by Eq.~(\ref{chi_int-mon-dim}) when simultaneously taking into consideration the validity of Eqs. (\ref{hef_a}) and (\ref{j_eff}) we have found for the magnetic compound  Cu$_3$(P$_2$O$_6$OH)$_2$ the following set of the microscopic parameters  $g=2.2$, $J_1/k_B=102.67$~K, $J_2/k_B=29.83$~K. According to this fitting set, the intermediate 1/3-plateau should appear in the magnetic-field range between 10.14~T and 79~T. 
Note furthermore that two different sets of the coupling constants were previously suggested for the magnetic compound  Cu$_3$(P$_2$O$_6$OH)$_2$, namely, $J_1/k_B=98$~K, $J_2/k_B=28$~K \cite{hase06} and $J_1=111/k_B$~K, $J_2/k_B=30$~K \cite{hase07,hase08}. To reach better quantitative  agreement with the low- and high-temperature susceptibility data of Cu$_3$(P$_2$O$_6$OH)$_2$ the developed strong-coupling method taking advantage of the concept of the effective coupling $J_{eff}$ has allowed us to refine both coupling constants.

\begin{figure}
\begin{center}
\includegraphics[width=0.48\textwidth]{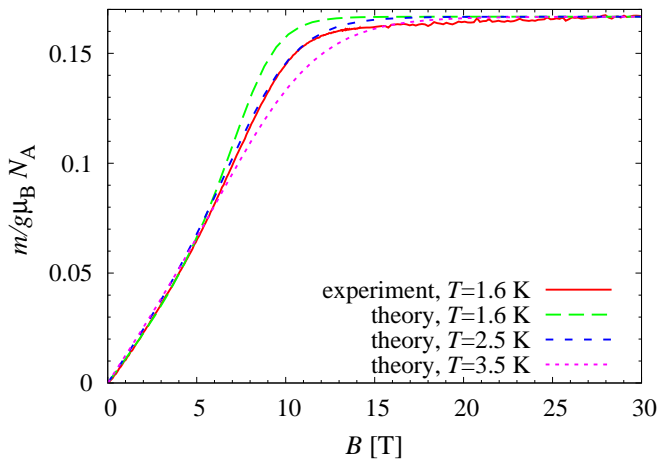}
\end{center}
\caption{(Color online) The isothermal magnetization curve of Cu$_3$(P$_2$O$_6$OH)$_2$ recorded at $T = 1.6$~K and the respective theoretical prediction made by the QMC simulations of the spin-$\frac{1}{2}$ Heisenberg trimerized chain with $L=120$ spins for the fitting set of the parameters: $J_1/k_B=102.67$~K, $J_2/k_B=29.83$~K, $g=2.2$ ($5\times10^4$ QMC steps were used for thermalization and additional $5\times10^5$ QMC steps for statistical averaging). 
}
\label{fig:mag_real}
\end{figure}

\begin{figure}
\begin{center}
\includegraphics[width=0.48\textwidth]{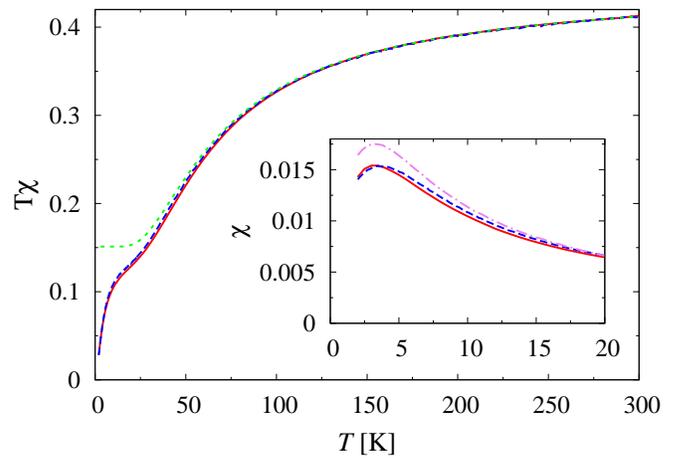}
\end{center}
\caption{(Color online) The temperature dependence of the susceptibility times temperature product for Cu$_3$(P$_2$O$_6$OH)$_2$. The experimental data (solid lines) are compared with the theoretical data (dashed lines), which were obtained  by the QMC simulations of the spin-$\frac{1}{2}$ Heisenberg trimerized chain with $L=120$ spins for the fitting set of the parameters: $J_1/k_B=102.67$~K, $J_2/k_B=29.83$~K, $g=2.2$. A dotted line in the main panel is the result of the thermodynamic perturbation theory (\ref{chi_int-mon-dim}), while a dashed-dotted line in the inset displays the QMC data for the effective model (\ref{hef_a}).}
\label{fig:susc_real}
\end{figure}

The spin-$\frac{1}{2}$ Heisenberg trimerized as well as uniform chains are both amenable to the QMC simulations, whereas the relevant QMC data for the low-temperature magnetization curve and temperature dependence of the magnetic susceptibility are compared in Figs.~\ref{fig:mag_real} and \ref{fig:susc_real} with the relevant experimental data reported previously for Cu$_3$(P$_2$O$_6$OH)$_2$. It can be seen from Fig.~\ref{fig:susc_real} that the measured data for the magnetic susceptibility are closely followed in the whole temperature range by the theoretical results of the QMC simulations obtained for the found fitting set of the interaction parameters. On the other hand, it follows from Fig.~\ref{fig:mag_real} that the theoretical results for the magnetization coincide well with the experimental data only for sufficiently low magnetic fields, while they overestimate the experimental data at higher magnetic fields. One plausible explanation for this discrepancy is the adiabatic heating of the sample during the magnetization process, which can be supported by the theoretical magnetization data calculated at slightly higher temperatures. Finally, it is worthwhile to remark that the effective spin-$\frac{1}{2}$ Heisenberg chain (\ref{hef_a}) also predicts the low-temperature peak of the specific heat at $T\approx 2.44$~K, whereas similar feature has been also experimentally observed in Ref.~\onlinecite{hase06}.
 

\section{Conclusions}
\label{sec:concl}

To summarize, it has been shown that the strong-coupling approach can be substantially improved for the quantum Heisenberg chain when considering the perturbation about the exact solution of its Ising-Heisenberg counterpart. The improvement of the novel many-body perturbation scheme over the standard strong-coupling method closely relates to the fact that the unperturbed Hamiltonian of the Ising-Heisenberg model accounts for correlations between all interacting spins and it does not split the investigated quantum spin system into smaller noninteracting fragments. The paradigmatic example the spin-$\frac{1}{2}$ Heisenberg trimerized chain, for which the many-body perturbation theory was formulated about the exact solution of the spin-$\frac{1}{2}$ Ising-Heisenberg diamond chain, serves in evidence of this statement. As a matter of fact, the upper and lower critical fields of the intermediate one-third plateau of the spin-$\frac{1}{2}$ Heisenberg trimerized chain derived within the unconventional perturbation theory are in a perfect agreement with the numerical DMRG data up to relatively high values of the coupling ratio $J_2/J_1 \lesssim 0.5$ in contrast with the standard perturbation expansion developed from the monomer-dimer limit. In addition, we have also analyzed the applicability of the effective-model approach for an investigation of magnetization curves and thermodynamic properties at finite temperatures, whereas a plausible agreement with the numerical QMC data was observed at sufficiently low temperatures $k_BT/J_1\lesssim 0.2$. Besides, we have found that the effective description based on the modified perturbation theory reproduces well temperature position of the peaks experimentally detected in the susceptibility and specific heat data. However, the height of the susceptibility peak is slightly higher for the effective model than for the original  model. 

Last but not least, we have suggested the relatively simple procedure to determine the microscopic parameters for the magnetic compounds with strong intradimer and weak monomer-dimer coupling, which provide suitable experimental realizations of the spin-$\frac{1}{2}$ Heisenberg trimerized chain. The idea how to unambiguously determine both coupling constants of the spin-$\frac{1}{2}$ Heisenberg trimerized chains is based on the combination of the low- and high-temperature behavior of the magnetic susceptibility. The applied procedure gave for the prototype of the trimerized spin chain Cu$_3$(P$_2$O$_6$OH)$_2$ the following set of the microscopic parameters $J_1/k_B=102.67$~K, $J_2/k_B=29.83$~K, $g=2.2$, which refines the parameter sets reported in the previous studies \cite{hase06,hase07,hase08}. However, the suggested procedure can be also applied for an appropriate description of magnetic and thermodynamic properties of other polymeric compounds, which have a magnetic structure of the spin-$\frac{1}{2}$ Heisenberg trimerized chain as for instance copper-based coordination polymers Cu$_3$(OH)$_4$SO$_4$ \cite{vilminot03,vilminot07}, [Cu$_3$Cl$_4$(C$_4$H$_9$NO$_2$)$_4$(H$_2$O)$_2$](ClO$_4$)$_2$ \cite{bouh17} and [Cu$_3$(acetate)$_6$(imidazole)$_2$] \cite{machado18}. 

It is our hope that the present article opens up further possibilities for a more accurate investigation of other fully quantum Heisenberg spin models on the grounds of the unconventional perturbation theory developed from exactly solved Ising-Heisenberg spin systems (see Ref. \cite{strecka10} for methodological details and substantial survey of literature on the exactly solved Ising-Heisenberg models). For instance, the spin-$\frac{1}{2}$ Heisenberg tetrameric chain with a regular alternation of two ferromagnetic and two antiferromagnetic exchange couplings as a theoretical model of the polymeric compound Cu(3-Chloropyridine)$_2$(N$_3$)$_2$ represents another valuable example of a quantum spin chain, where the perturbation expansion could be straightforwardly derived from the respective exact solution of a tetramer Ising-Heisenberg bond alternating chain \cite{strecka05}. Moreover, it turns out that the nature of unconventional fractional magnetization plateaus of two-dimensional Shastry-Sutherland model as one of the most challenging and widely discussed problems of modern quantum magnetism can also be interpreted within the novel many-body perturbation theory developed from the exact solution for a spin-$\frac{1}{2}$ Ising-Heisenberg model on the Shastry-Sutherland lattice \cite{verkholyak14,verkholyak21}.

\begin{acknowledgments}
The authors are grateful to Prof. Hase for providing the experimental data for the magnetization and susceptibility data of the polymeric compound Cu$_3$(P$_2$O$_6$OH)$_2$. The authors would like also to thank Prof. Derzhko and Dr. Krupnitska for the useful comments and remarks on the manuscript. T.V. acknowledges the financial support provided by the National Scholarship Programme of the Slovak Republic for the Support of Mobility of Students, PhD Students, University Teachers, Researchers and Artists. J.S. acknowledges financial support provided by Slovak Research and Development Agency provided under the contract No. APVV-16-0186 and by The Ministry of Education, Science, Research and Sport of the Slovak Republic provided under the grant No. VEGA 1/0105/20.
\end{acknowledgments}

\appendix

\section{Exact solution of the Ising-Heisenberg diamond chain}
\label{app:IH-dc}
Let us utilize the method based on the projection operators in order to find the exact solution for the spin-$\frac{1}{2}$ Ising-Heisenberg diamond chain given by the Hamiltonian (\ref{ham_IH}),  which can be decomposed into the sum of local Hamiltonians $H_{IH}=\sum_{i} H^{(0)}_i$ each of them related to a four-spin diamond cluster:  
\begin{eqnarray}
\label{IH-cluster}
H^{(0)}_{i}&=&J_1{\mathbf s}_{1,i}\cdot{\mathbf s}_{2,i} + \frac{J^z_2}{2}(s^z_{1,i} + s^z_{2,i})(s^z_{3,i-1}+s^z_{3,i})
\nonumber\\
&&-h\left[s^z_{1,i} + s^z_{2,i} + \frac{1}{2}(s^z_{3,i-1}+ s^z_{3,i})\right].
\end{eqnarray}
As the first step, it is advisable to introduce the dimer-states basis for strongly coupled spin pairs:
\begin{eqnarray}
&&|0\rangle_i=\frac{1}{\sqrt2}(|\!\uparrow\rangle_{1,i}|\!\downarrow\rangle_{2,i}-|\!\downarrow\rangle_{1,i}|\!\uparrow\rangle_{2,i}),
\nonumber\\
&&|1\rangle_i=|\!\uparrow\rangle_{1,i}|\!\uparrow\rangle_{2,i},
\nonumber\\
&&|2\rangle_i=\frac{1}{\sqrt2}(|\!\uparrow\rangle_{1,i}|\!\downarrow\rangle_{2,i}+|\!\downarrow\rangle_{1,i}|\!\uparrow\rangle_{2,i}),
\nonumber\\
&&|3\rangle_i=|\!\downarrow\rangle_{1,i}|\!\downarrow\rangle_{2,i},
\label{dimerbasis}
\end{eqnarray}
and the respective projection operators on these dimer states \cite{parkinson79}:
\begin{eqnarray}
A^{ab}_i=|a\rangle_i\langle b|_i.
\label{proj-operators}
\end{eqnarray}
The correspondence between the spin and projection operators can be found by a straightforward calculation (see  for instance Ref.~\onlinecite{parkinson79}). The four-spin diamond cluster Hamiltonian \eqref{IH-cluster} becomes diagonal in terms of new projection operators as it can be rewritten as follows:
\begin{eqnarray}
\label{IH-cluster_a}
H^{(0)}_{i}&=&J_1\left(\frac{1}{4}-A_i^{00}\right) 
+ \left[\frac{J^z_2}{2}(s^z_{3,i-1}+s^z_{3,i})-h\right]
\nonumber\\
&&\times (A^{11}_{i} - A^{33}_{i}) 
-\frac{h}{2}(s^z_{3,i-1}+ s^z_{3,i}).
\end{eqnarray}
Owing to this fact, one can perform the decoration-iteration transformation for the dimer spins in order to obtain the statistical mechanics of the spin-$\frac{1}{2}$ Ising-Heisenberg diamond chain quite rigorously  \cite{canova06}. In the present paper we are mostly interested in the ground-state properties. For this purpose, it is necessary to find eigenenergies of the four-spin cluster Hamiltonian \eqref{IH-cluster}, whose two lowest possible eigenvalues are given in the notation $E^{(0)}_i(s^z_{3,i-1},l_i,s^z_{3,i})$:
\begin{eqnarray}
E^{(0)}_i(\uparrow,0,\uparrow)&=&-\frac{3 J_1}{4}-\frac{1}{2} h,
\nonumber\\
E^{(0)}_i(\uparrow,1,\uparrow)&=&\frac{J_1}{4}+\frac{J^z_2}{2}-\frac{3}{2} h,
\end{eqnarray}
where $l_i$ is the index pertinent to the eigenstates of the $i$-th dimer ($l_i = 0$ is assigned to the singlet state and $l_i = 1$ to the polarized triplet state). If the local state with the minimal energy is known, we are able to find the corresponding ground state by extending the relevant eigenstate to the whole system. Since the coupling constant $J_2$ is supposed to be much smaller than  the one $J_1$, we limit ourselves only to the part of the ground-state phase diagram in the parameter space $J_2<J_1$. At sufficiently low magnetic fields $0 \leq h < h_c \equiv J_1+J^z_2/2$ the ground state is the MD phase described by the following eigenvector:
\begin{eqnarray}
\label{MD}
|{\rm MD}\rangle=\prod_{i=1}^N |0\rangle_i|s^z\rangle_{3,i}.
\end{eqnarray}
The monomer-dimer state \eqref{MD} is macroscopically degenerate in zero field, since the monomeric spins are completely free (paramagnetic) as they are isolated through non-magnetic singlet dimers $|0\rangle_i$. However, this ground-state degeneracy is completely lifted by any nonzero magnetic field being responsible for a full polarization of the monomeric spins ${\mathbf s}_{3,i}$. At high enough magnetic fields $h>h_c$ the ground state corresponds to the SAT phase with all spins being fully aligned to the magnetic-field direction given by the eigenvector: 
\begin{eqnarray}
\label{SAT}
|{\rm SAT}\rangle=\prod_{i=1}^N |1\rangle_i|\uparrow\rangle_{3,i}.
\end{eqnarray}
The spin-$\frac{1}{2}$ Ising-Heisenberg diamond chain also becomes macroscopically degenerate due to an energy equivalence of the singlet state $|0\rangle_i$ and the polarized triplet state $|1\rangle_i$ at the saturation field $h = h_c$, which determines the phase boundary between the MD and SAT ground states.  

\section{Perturbation theory for small magnetic fields}
\label{app:small_fields}

In this part we will develop the perturbation theory applicable for the spin-$\frac{1}{2}$ Heisenberg trimerized chain  at sufficiently small magnetic fields. To this end, one should take for the unperturbed spin-$\frac{1}{2}$ Ising-Heisenberg diamond-chain model (\ref{ham_IH}) the following value of the field term $h'=0$. For further convenience, the perturbed part of the Hamiltonian can be decomposed into a sum of the local terms $V=\sum_{i=1}^N (V_{i} + V_{i,i+1})$, which can be explicitly written in terms of the projection operators as:
\begin{eqnarray}
\label{pert_a}
V_{i,i+1}&=&\frac{J^z_2}{2}(A_i^{20}+A_i^{02})(s^z_{3,i-1}-s^z_{3,i})
\nonumber\\
&&+\frac{J^{xy}_2}{2}(s^+_{3,i-1}s^-_{1,i} + s^+_{2,i}s^-_{3,i} + {\rm h.c.}),
\nonumber\\
V_{i}&=&-h(A_i^{11}-A_i^{33}+s^z_{3,i}).
\end{eqnarray}
The projection on the ground state can be constructed as $P_0=\prod_i A^{00}_i$. The first-order contribution to the effective Hamiltonian is as follows:
\begin{eqnarray}
\label{h_eff1}
H^{(1)}_{eff}=P_0 V P_0=-h \sum_{i}s^z_{3,i}.
\end{eqnarray}
To obtain the perturbation terms up to the second order, one needs to calculate first the action of the perturbation term on the projection operator
\begin{eqnarray}
V_{i,i+1} P_0&=&
\bigg\{
\frac{J^z_2}{2}A_i^{20}(s^z_{3,i-1}{-}s^z_{3,i})
{+}\frac{J^{xy}_2}{2\sqrt2}\left[A_i^{30}(s^+_{3,i-1}{-}s^+_{3,i}) 
\right.
\nonumber\\ &&
\left. 
{+} A_i^{10}(s^-_{3,i} {-} s^-_{3,i-1}) \right] {-} hs^z_{3,i}
\bigg\}P_0, 
\end{eqnarray}
as well as the excitation energies $\Delta E(s^z_{i-1},l_i,s^z_{i})=E(s^z_{i-1},l_i,s^z_{i})- E(s^z_{i-1},0,s^z_{i})$ for a change of the singlet state ($l_i=0$) of the $i$-th dimer to any triplet state ($l_i=1,2,3$):
\begin{eqnarray}
\Delta E(s^z_{i-1},1,s^z_{i})&=&J_1+\frac{J^{z}_2}{2}(s^z_{i-1}+s^z_{i}),
\nonumber\\
\Delta E(s^z_{i-1},2,s^z_{i})&=&J_1,
\nonumber\\
\Delta E(s^z_{i-1},3,s^z_{i})&=&J_1-\frac{J^{z}_2}{2}(s^z_{i-1}+s^z_{i}).
\label{dE}
\end{eqnarray}
Inserting these results into Eq.~(\ref{eq_pert}) and assuming the isotropic limit $J_2^{xy} = J_2^z = J_2$ one gets the second-order perturbation term:
\begin{eqnarray}
\label{hef_a2}
H_{eff}&=&J_{eff}\sum_{i=1}^{N} {\mathbf s}_{3,i}\cdot{\mathbf s}_{3,i+1}, 
\nonumber\\
J_{eff}&=&\frac{J_2^2}{2J_1-J_2}>0,
\end{eqnarray}
which gives after complementing the first-order contribution \eqref{h_eff1} the effective Hamiltonian (\ref{hef_a}) considered in Sec.~\ref{subs:small_fields}.

\section{Perturbation theory for high magnetic fields}
\label{app:high_fields}
The perturbation theory for high magnetic fields is based on the expansion near the saturation field $h_c = J_1+J^z_2/2$, where the magnetization of the unperturbed spin-$\frac{1}{2}$ Ising-Heisenberg diamond-chain model (\ref{ham_IH}) exhibits a discontinuous jump from $1/3$-plateau to the saturation value. The projection operator for this degenerate state can be written as 
$P_0=\prod_{i}(A^{00}_i+A^{11}_i)|\!\uparrow\rangle_{3,i}\langle\uparrow\!|_{3,i}$. The first order term can be easily calculated as
\begin{eqnarray}
\label{Hh_eff1}
H^{(1)}_{eff}=P_0 V P_0=-(h-h_c) \sum_{i} \left(A^{11}_i+\frac{1}{2}\right).
\end{eqnarray}
To get the second-order perturbation term it is necessary to calculate first the action of the perturbed term on the projection operator:
\begin{eqnarray}
V_{i,i+1} P_0&=&\left\{\frac{J^{z}_2}{4}({-}A^{20}_i {+} A^{20}_{i+1}) 
{-}\frac{J^{xy}_2}{2\sqrt{2}}(A^{10}_i {+} A^{10}_{i+1}) s^{-}_{3,i} 
\right. \nonumber\\ &&\left.
-(h-h_c)\left(A^{11}_i+\frac{1}{2}\right)
\right\}P_0,
\end{eqnarray}
which gives after compiling all calculations the following result for the second-order term:
\begin{eqnarray}
\label{Hh_eff2}
H^{(2)}_{eff}&=&\frac{(J^{xy}_2)^2}{4}\sum_{i} P_0\left\{
{-}\frac{A^{00}_i A^{00}_{i+1}}{J_1} {-} \frac{A^{00}_i A^{11}_{i+1} {+} A^{11}_i A^{00}_{i+1}}{2J_1-J^{z}_2}
\right. \nonumber\\ &&\left.
+\frac{A^{01}_i A^{10}_{i+1} + A^{10}_i A^{01}_{i+1}}{2J_1-J^{z}_2}
\right\}P_0.
\end{eqnarray}
Next, let us introduce the pseudo-spin formalism identifying the singlet (triplet) state of the dimer with $|\tilde\downarrow\rangle_{i}$ ($|\tilde\uparrow\rangle_{i}$), i.e. $A^{00}_i=\frac{1}{2} - \tilde{s}^z_i$, $A^{11}_i=\frac{1}{2} + \tilde{s}^z_i$, $A^{01}_i=\tilde{s}^{-}_i$, $A^{10}_i=\tilde{s}^{+}_i$. Imposing $J_2^{xy} = J_2^z = J_2$, the second-order contribution (\ref{Hh_eff2}) to the effective Hamiltonian can be rewritten in terms of the pseudospin operators as follows:
\begin{eqnarray}
\label{Hh_eff2s}
H^{(2)}_{eff}&=&\sum_i \left[ 
J^{xy}_{eff}(\tilde{s}^x_i \tilde{s}^x_{i+1} {+} \tilde{s}^y_i \tilde{s}^y_{i+1})
{+} J^z_{eff}\tilde{s}^z_i \tilde{s}^z_{i+1}
{+} h_2 \tilde{s}^z_i  
\right],
\nonumber\\
J^{xy}_{eff}&=&\frac{J_2^2}{ 4J_1\left(1 {-} \frac{J_2}{2J_1}\right)}>0,
\nonumber\\
J^z_{eff}&=&\frac{J_2}{2J_1}J^{xy}_{eff}>0,
\nonumber\\
h_2&=&\frac{(J^{xy}_2)^2}{4J_1}.
\end{eqnarray}
The first-order \eqref{Hh_eff1} and second-order \eqref{Hh_eff2s} terms are summed up in the effective Hamiltonian given by Eq.~(\ref{hef_b}).

\end{document}